# On resampling methods for model assessment in penalized and unpenalized logistic regression

Angelika Geroldinger [1], Lara Lusa [2,3], Mariana Nold [4], and Georg Heinze [5]

[1] Center for Medical Statistics, Informatics and Intelligent Systems, Medical University of Vienna, Spitalgasse 23, 1090 Vienna, Austria; angelika.geroldinger@meduniwien.ac.at
[2] Faculty of Mathematics, Natural Sciences and Information technologies, University of Primorska, Glagoljaška 8, SI-6000 Koper, Slovenia; lara.lusa@famnit.upr.si
[3] Faculty of Medicine, University of Ljubljana, Vrazov trg 2, 1000 Ljubljana, Slovenia; lara.lusa@mf.uni-lj.si
[4] Department of Sociology, Friedrich Schiller University Jena, Carl-Zeiß-Straße 3, 07743 Jena, Germany; mariana.nold@uni-jena.de
[5] Center for Medical Statistics, Informatics and Intelligent Systems, Medical University of Vienna, Spitalgasse 23, 1090 Vienna, Austria; georg.heinze@meduniwien.ac.at

**Abstract:** Penalized logistic regression methods are frequently used to investigate the relationship between a binary outcome and a set of explanatory variables. The model performance can be assessed by measures such as the concordance statistic (c-statistic), the discrimination slope and the Brier score. Often, data resampling techniques, e.g. crossvalidation, are employed to correct for optimism in these model performance criteria. Especially with small samples or a rare binary outcome variable, leave-one-out crossvalidation is a popular choice. Using simulations and a real data example, we compared the effect of different resampling techniques on the estimation of c-statistics, discrimination slopes and Brier scores for three estimators of logistic regression models, including the maximum likelihood and two maximum penalized-likelihood estimators. Our simulation study confirms earlier studies reporting that leave-one-out crossvalidated c-statistics can be strongly biased towards zero. In addition, our study reveals that this bias is more pronounced for estimators shrinking predicted probabilities towards the observed event rate, such as ridge regression. Leave-one-out crossvalidation also provided pessimistic estimates of the discrimination slope but nearly unbiased estimates of the Brier score. We recommend to use leave-pair-out crossvalidation, five-fold crossvalidation with repetition, the enhanced or the .632+ bootstrap to estimate c-statistics and leave-pair-out or five-fold crossvalidation to estimate discrimination slopes.

## 1. Introduction

The concordance statistic (c-statistic) is a widely used measure to quantify the discrimination ability of models for binary outcome variables. Calculating the c-statistic for the data on which the model was fitted will usually give too optimistic results, especially with small samples or rare events. Attempting to correct for this over-optimism, data resampling techniques such as crossvalidation (CV) or the bootstrap are frequently employed. Leave-one-out (LOO) CV has the advantage of being applicable even with small samples where other techniques such as ten-fold or five-fold CV might run into problems. With LOO CV only the *pooling* strategy can be used to estimate the c-statistic: the crossvalidated probabilities, each derived from a different model, are eventually pooled to calculate a single c-statistic. With five-fold or ten-fold CV we usually apply an *averaging* strategy: the crossvalidated probabilities of the observations included in each left-out fold are used to evaluate the statistics of interest and the final crossvalidated statistics are obtained by averaging the statistics computed in each fold. Only the averaging approach is a proper CV as it evaluates the statistic of interest in each left-out fold. Whereas for statistics applicable to single observations such as the Brier score, LOO CV is known to yield nearly unbiased estimates [1], it was shown that for the c-statistic LOO CV can result in severe bias towards 0 [2,3].

The discrimination slope [4] is an increasingly popular measure of predictive accuracy in binary models. Its construction parallels that of the c-statistics in some aspects but it is unclear if it suffers from similar problems with LOO CV. Furthermore, it is unknown whether the magnitude of this bias is similar when comparing different penalized-likelihood estimators for logistic regression models. Therefore, we studied the bias and variance in c-statistics and discrimination slopes, combining several logistic regression penalized-likelihood estimators with several resampling techniques in a simulation study with factorial design. We also considered the Brier score in our simulation study as a measure applicable to single observations, where we expect LOO CV to give unbiased estimates [1].

We focus on the situation of relatively few observations or events, where reliable predictive models are perhaps out of scope. Instead, we assume that the aim of the data analysis is to summarize or represent the data structure in a compact manner, which was termed descriptive modeling [5]. Not only in predictive but also in descriptive modeling should the assessment of the discriminatory ability be part of the model building process.

The remainder of this paper is organized as follows: first, we explain the resampling techniques and model estimators of interest. A study on the association between diabetes and the waist-hip ratio serves as illustrative example. Subsequently, we provide an intuitive explanation of the problems with LOO CV using simply structured artificial data. Next, design and results of our simulation study are described. Finally, we discuss the impact of our findings on routine statistical analyses.

## 2. Methods

*2.1. Measures of model performance*

We denote the two outcome values as 'event' and 'non-event', and assume that in logistic regression the probability of an event is modeled.

The **c-statistic** is the proportion of pairs among all possible pairs of observations with contrary outcomes in which the predicted event probability is higher in the observation with the event than in the observation with the non-event. It equals the area under the receiver operating characteristic curve.

The **discrimination slope** is the difference between the mean predicted probability of an event for observations with events and the mean predicted probability for observations with non-events. Paralleling the construction of the c-statistic, the discrimination slope can be computed as the average pairwise difference in predicted probabilities, thus representing a parametric version of the c-statistic. It was suggested as 'highly recommendable $R^2$-substitute for logistic regression models' by Tjur [4] and recently revisited by Antolini et al [6].

The **Brier score** is the mean squared difference between the binary outcome and the predicted probabilities [7]. It equals 0 for perfect models. The magnitude of the Brier score has to be interpreted in the context of the properties of the data set. For instance, for a non-informative model (with predicted probabilities equal to the event rate) the Brier score equals 0.25 for an event rate of 0.5 but equals 0.1875 for an event rate of 0.25. Unlike the c-statistic and the discrimination slope, the Brier score can technically be computed for single observations.

*2.2. Techniques to correct for over-optimism*

Next, we describe some resampling techniques by means of computation of optimism-corrected estimates of the c-statistic. If not mentioned otherwise, methodology straightforwardly generalizes to the discrimination slope and the Brier score. We denote by '**apparent**' measures that are calculated from the data on which the model was fitted.

With $f$**-fold CV** the data are split into $f$ approximately equally sized parts, often called the 'folds'. A model is fitted on the observations contained in $f$-1 parts of the data. Using this model, predicted probabilities for the observations in the excluded $f$-th part are calculated. By excluding each fold in turn, one obtains $f$ c-statistics which are then averaged. To decrease variability caused by the random partition into folds, the whole procedure is repeated $r$ times and results are averaged. Here we consider $f = 5$ and $r = 40$, i.e. 5-fold CV with 40 repetitions. As we always performed 5-fold CV repeatedly, we will often omit the specification 'with 40 repetitions' for the sake of brevity.

Setting $f$ to the sample size $n$ ('**leave-one-out CV**'), the c-statistic cannot be computed for the excluded part as it contains only one observation. Instead, one fits $n$ models for all possible subsets of $n-1$ observations, calculates the predictive probabilities for the respective left-out observations and computes a single c-statistic from the pooled $n$ predicted probabilities.

Another approach that is independent of random sampling but based on c-statistics calculated within folds is **leave-pair-out (LPO) CV** [2,3]. With LPO CV each pair of observations with contrary outcomes is omitted in turn from the data, a model is fitted on the remaining $n-2$ observations and predicted probabilities for the two excluded observations are calculated from this model. The LPO crossvalidated c-statistic is the proportion of pairs where the predicted probability of the observation with the event is higher than that of the observation with the non-event. LPO CV can imply considerable computational burden: if $k$ is the number of events, $(n-k)k$ models have to be estimated, compared to only $n$ models with LOO CV. For example, with 50 events among 100 observations, 2 500 models must be fitted with LPO but only 100 with LOO CV. Whereas LPO CV generalizes straightforwardly to the discrimination slope it is not clear how it should be adapted for the Brier score: Simply averaging the Brier score computed for all left-out pairs will give biased estimates in the case of unbalanced outcomes because of the dependence of the Brier score on the event rate. One solution would be to adequately weight contributions by events and non-events. Here we refrain from applying the LPO CV to the Brier score.

With the **enhanced bootstrap** [8], the bias due to overfitting is explicitly estimated and then subtracted from the apparent c-statistic. Specifically, 200 samples of $n$ observations with replacement are drawn from the original data set. On each of these bootstrap resamples a model is fitted and used to calculate c-statistics both for the bootstrap resample and the original data. An estimate of 'optimism' is obtained by subtracting the average c-statistic in the original data from the average c-statistic in the bootstrap resamples. The enhanced bootstrap c-statistic is then given by the apparent c-statistic minus the estimate of optimism.

The **.632+ bootstrap** [9] is a weighted average of the apparent c-statistic and the average 'out-of-the-bag' c-statistic calculated from bootstrap resamples. The 'out-of-the-bag' c-statistic is obtained by fitting the model in a bootstrap resample and applying it to the observations not contained in that bootstrap resample. We give the technical details in the Appendix.

*2.3. Penalized-likelihood estimation methods*

We investigated the performance of the resampling techniques in combination with the following estimators of logistic regression:
- maximum likelihood estimation (ML),
- Firth's penalized logistic regression (FL) [10,11],
- logistic ridge regression (RR) [12].

FL amounts to penalization by the Jeffreys prior and was shown to reduce the bias in coefficient estimates compared to ML. With RR, the log-likelihood is penalized by the square of the Euclidean norm of the regression parameters multiplied by a tuning parameter. We chose the tuning parameter by minimizing a penalized version of the Akaike's Information Criterion AIC given by $-2l(\hat{\beta}) + 2\, df_e$ with $l(\hat{\beta})$ the log-likelihood,

$$df_e = \text{trace}\left(\frac{\partial^2 l}{\partial \beta^2}(\hat{\beta})\left(\frac{\partial^2 l^*}{\partial \beta^2}(\hat{\beta})\right)^{-1}\right)$$

the effective degrees of freedom and $l^*(\hat{\beta})$ the penalized log-likelihood [13]. This approach of optimizing the tuning parameter is less computer-intensive than the optimization of crossvalidated measures and has been reported to yield similar or even superior results [8].

For the implementation of ML, and FL we used the R-package logistf with default convergence criteria. For RR we applied the function lrm in the R-package rms, setting the singularity criterion to $10^{-15}$.

If the data are separated, i.e. if a combination of explanatory variables perfectly predicts the outcome, then ML fails to produce finite regression coefficients and some predicted probabilities will be exactly 0 or 1 [14]. By contrast, FL gives reasonable results in the case of separation. Under separation, RR will supply finite regression coefficients if the tuning parameter is greater than 0. However, CV or AIC optimization will often set the tuning parameter to 0 in case of separation, and then RR leads to the same problems as ML [14]. See section S1 for how we handled separation, linearly dependent explanatory variables or binary outcomes restricted to one category occurring in bootstrap resamples or CV subsets.

**3. Motivation**

*3.1. A real data example: association between waist-hip ratio and diabetes*

As part of a smoking cessation project among blacks in two rural Virginia counties, a screening examination on coronary heart disease risk factors was performed [15]. For illustration purposes, we focus on the association between the waist-hip ratio and the presence of diabetes (defined by glycosylated hemoglobin $> 7.0$), adjusted for gender, in the Virginia county Louisa. Among the 198 study participants, 14.6% (29 persons) were classified as having diabetes. With ML, an increase of 0.10 in the waist-hip ratio was associated with an adjusted odds ratio of 1.9 (95% CI; 1.01, 3.58). On the level of predicted probabilities, this can be expressed as a probability of diabetes of 0.112 versus 0.193 for females with a waist-hip ratio of 0.8 versus 0.9. In line with findings from previous studies [2,3], LOO CV resulted in lower c-statistics for ML, FL and RR than the other resampling techniques, see Figure 1. A model with no discriminative ability (a 'random guess') would yield a c-statistic of approximately 0.5. In this respect, it is remarkable that LOO crossvalidated c-statistics of RR was 0.468 (smaller than 0.5) while the corresponding c-statistics of ML and FL were 0.54 (larger than 0.5), which would lead to the possibly misleading conclusion that RR supplies a model that performs worse than a random guess, while ML and FL yield better models. All other resampling techniques gave similar c-statistics across model estimators.

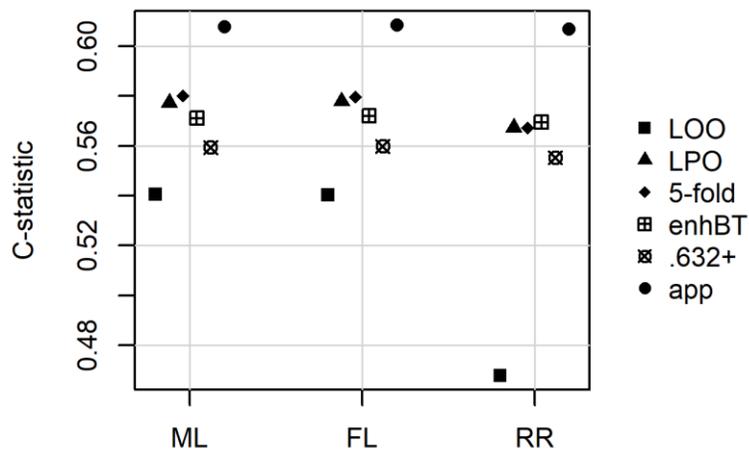

**Figure 1.** Apparent and optimism-corrected c-statistics for three different estimators of logistic regression models using data from a screening examination on coronary heart disease risk factors.
ML, maximum likelihood; FL, Firth's logistic regression; RR, ridge regression.
LOO, leave-one-out crossvalidation; LPO, leave-pair-out crossvalidation; 5-fold, 5-fold crossvalidation; enhBT, enhanced bootstrap; .632+, .632+ bootstrap; app, apparent estimate.

*3.2. The bias in LOO crossvalidated c-statistics*

Figure 2 explains the bias in LOO crossvalidated c-statistics by illustrating the estimation process on simply structured, artificial data. The crucial observation in Figure 2 is that the predicted probability for the left-out observation was on average lower for events (CV cycles 1-5) than for non-events (CV cycles 6-20). This is not surprising: if an event was left out, the data used in the model fitting consisted of only 4 events out of 19 observations, compared to 5 events out of 19 observations if a non-event was left out. These considerations explain why LOO crossvalidated c-statistics (and discrimination slopes) are downward-biased; they are obtained by pooling the crossvalidated predicted probabilities derived from models estimated on data subsets with different event rates. Furthermore, Figure 2 illustrates that the bias in LOO crossvalidated c-statistics usually is more severe for models yielding predicted probabilities with lower variance such as ridge regression. This tendency can lead to undesired results if one optimizes the tuning parameter in RR using LOO crossvalidated c-statistics, see Figure S1. Whereas for the null scenario the discrimination ability of RR is independent of the penalization strength, optimization of LOO crossvalidated c-statistics favors models with less regularization.

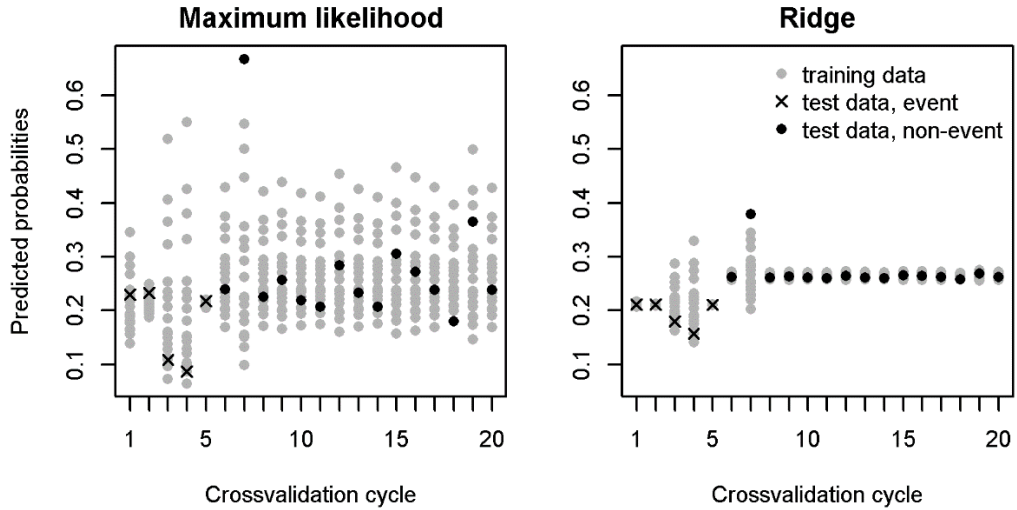

**Figure 2.** Illustration of the calculation of the c-statistic in leave-one-out crossvalidation for the maximum likelihood and ridge estimators. Data consisted of 20 observations of a normally distributed explanatory variable, with 5 randomly chosen observations labelled as 'events' (t-test p-value=0.584). Each tick on the x-axis corresponds to one of the 20 cycles in leave-one-out crossvalidation. Grey symbols mark predicted probabilities for the 19 observations used in the model fitting, black symbols mark the predicted probabilities for the left-out observations. Black crosses or circles indicate that the left-out observation corresponds to an event or non-event, respectively. The leave-one-out crossvalidated c-statistic was equal to 0.17 for maximum likelihood estimation and equal to 0 for ridge regression.

**4. Simulation study**

We follow the ADEMP structured approach in describing the setup of our simulation study [16].

*4.1. Aim*

The aim of the simulation study was to compare the accuracy of the resampling techniques LOO CV, LPO CV, 5-fold CV, enhanced bootstrap and .632+ bootstrap in estimating c-statistics, discrimination slopes and Brier scores for the model estimators ML, FL and RR.

*4.2. Data generating mechanism*

Data generation was motivated by the structure of real data sets, where typically a mix of variables with different distributions is encountered [17]. By sampling from a multivariate normal distribution and applying certain transformations, we generated one binary, one ordinal and three continuous explanatory variables, see section S2 for details. Binary outcomes $y_i$ were drawn from Bernoulli distributions with the event probability following a logistic model. We considered twelve simulation scenarios in a factorial design combining sample size ($n \in \{50,100\}$), marginal event rate ($E(y) \in \{0.25, 0.5\}$) and effect size (strong or weak effects of all explanatory variables, or null scenarios with no effects). More information on the magnitude of the effects is given in section S2. For each scenario we created 1 000 data sets.

*4.3. Estimands*

Our estimands are the c-statistic, the discrimination slope and the Brier score for the model estimators ML, FL and RR.

*4.4 Methods*

For each simulated dataset and each model estimator we assessed the predictive accuracy in terms of c-statistics, discrimination slopes and Brier scores by the five resampling methods LOO CV, LPO CV, 5-fold CV, enhanced bootstrap and .632+ bootstrap. The process mimicked the analysis of a real study where an external validation set is not available.

*4.5. Performance measures*

We compared the resampling-based c-statistics, discrimination slopes and Brier scores with those obtained if the estimated models were validated in the population, in our study approximated by an independent validation data set consisting of 100 000 observations. We described the performance of the resampling techniques in terms of mean and root mean squared difference (RMSD) of the resampling-based c-statistics, discrimination slopes and Brier scores to their respective independently validated (IV) counterparts. Finally, we calculated Monte Carlo standard errors for the mean squared difference [16] and the RMSD [18].

*4.6. Results*

First, we describe the distribution of the c-statistic, discrimination slope and Brier score obtained in the independent validation set, which will serve as gold standard. The mean IV c-statistics ranged between 0.5 and 0.684, see Table S2. RR achieved the largest mean IV c-statistics in non-null scenarios, but there was little difference between model estimators.

For the mean IV discrimination slope, the differences between the model estimators showed a range of up to 0.04 units, see Table S3. In non-null scenarios, ML achieved the largest median IV discrimination slopes, with values of up to 0.135. RR yielded the smallest median IV discrimination slopes, which were at least 20% smaller than by ML in all scenarios.

Results for the IV Brier score were in contrast to those for the IV discrimination slope: now ML performed worst in all scenarios, while RR resulted in the smallest mean Brier scores in all but one scenarios, see Table S4.

In approximating IV c-statistics, LOO CV performed worst both with respect to mean difference (bias) and RMSD, see Figure 3. The downward bias was most severe for RR and amounted to -0.274 in the most unfavorable scenario. For this scenario, the bias with ML or FL was only about a quarter of the bias with RR. In all but two scenarios, the enhanced and the .632+ bootstraps yielded the smallest RMSD for RR, ML and FL. Notably, the RMSD increased with increasing effect size for the .632+ bootstrap, whereas it decreased for all other resampling methods as expected. This behavior can be understood by looking at the definition of the .632+ bootstrap, which restricts it to values greater than or equal to the minimum of the apparent c-statistic and 0.5, as reflected in a right-skewed distribution of the .632+ c-statistic especially for null scenarios. For all model estimators and all resampling techniques the RMSD decreased with increasing sample size and increasing event rate. The differences between resampling techniques were less pronounced with stronger effects, larger sample sizes and balanced event rate.

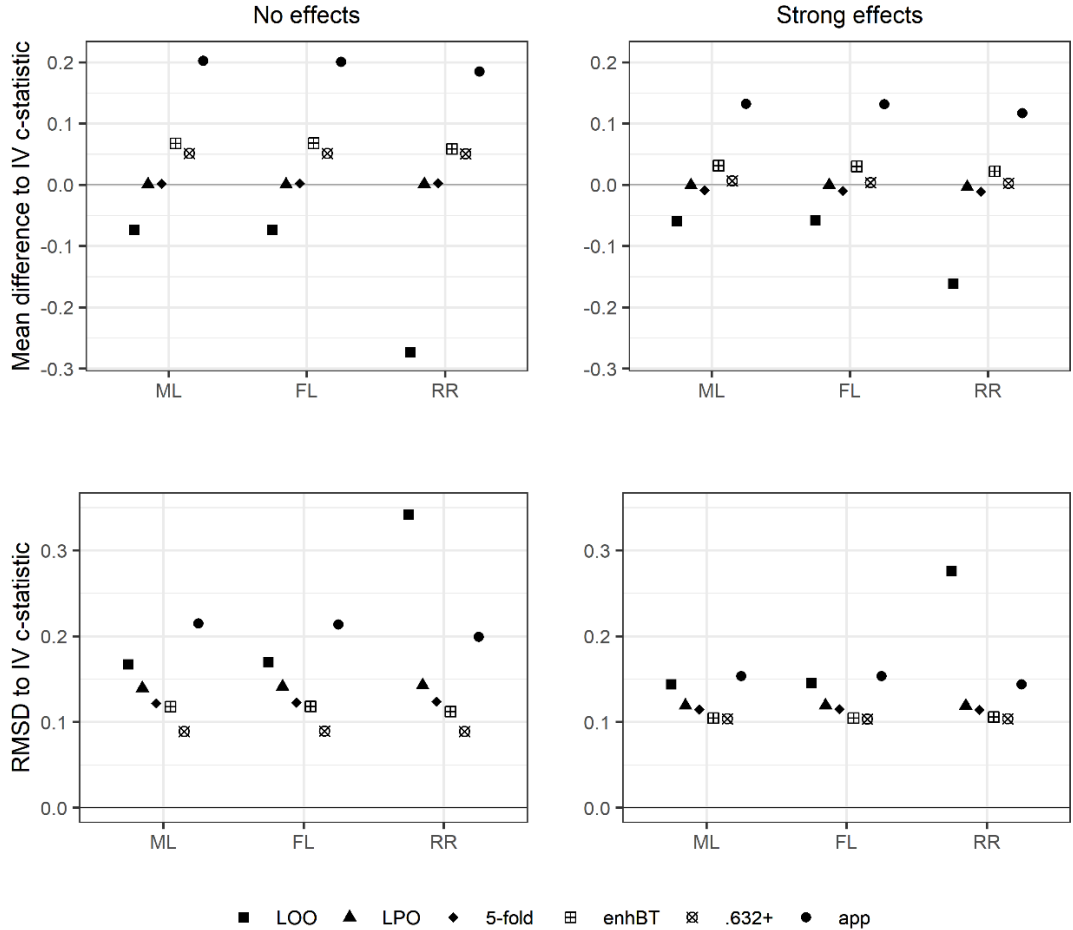

**Figure 3.** Mean differences and root mean squared differences (RMSD) between c-statistics computed by data resampling techniques and the independently validated (IV) c-statistic for the model estimators ML, FL and RR for the simulation settings with 50 observations, an event rate of 0.25 and either no or strong effects. The Monte Carlo standard errors of the mean difference and of the root mean squared difference (x100) were smaller than 0.8 and 0.7, respectively, for all scenarios.
ML, maximum likelihood; FL, Firth's logistic regression; RR, ridge regression.
LOO, leave-one-out crossvalidation; LPO, leave-pair-out crossvalidation; 5-fold, 5-fold crossvalidation; enhBT, enhanced bootstrap; .632+, .632+ bootstrap; app, apparent estimate.

LOO CV also performed poorly in approximating the IV discrimination slope, yielding pessimistic estimates with a RMSD at least larger than the one by LPO CV and 5-fold CV, see Figure 4. However, the differences in RMSD across resampling techniques were fairly small. Only for RR the two bootstrap techniques sometimes resulted in discrimination slopes with substantially larger RMSD than LPO CV and 5-fold CV. The .632+ bootstrap gave overly optimistic discrimination slopes, with an absolute mean difference to the IV values often larger than the one by LOO CV. On the other hand, in all but three simulation scenarios the .632+ bootstrap yielded discrimination slopes with smallest median deviations. This discrepancy can be explained by the right-skewness of the distribution of the differences between optimism-corrected and IV discrimination slopes, which was especially pronounced for the .632+ bootstrap. With increasing sample size, the RMSD decreased for all resampling techniques and all model estimators. Again, the differences between resampling techniques were less pronounced with increasing effect size, sample size and more balanced event rate.

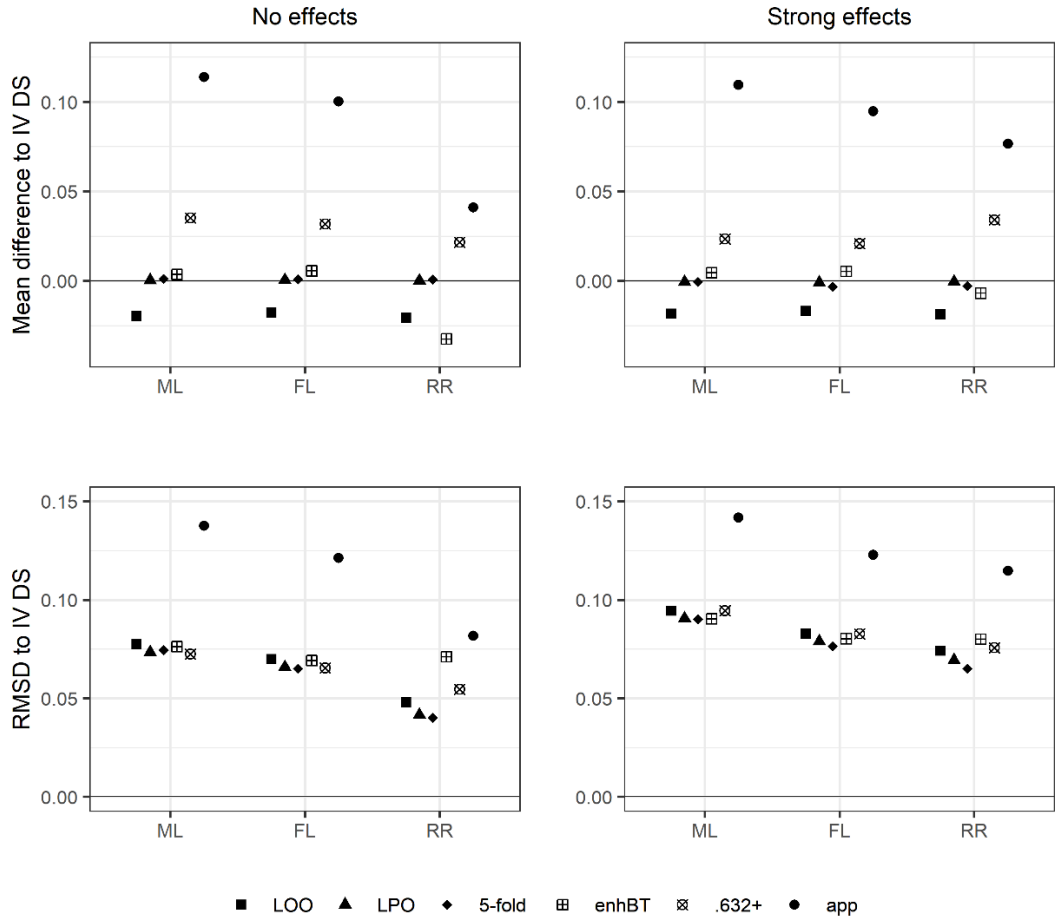

**Figure 4.** Mean differences and root mean squared differences (RMSD) between discrimination slopes (DS) computed by data resampling techniques and the independently validated (IV) DS for the model estimators ML, FL and RR for the simulation settings with 50 observations, an event rate of 0.25 and either no or strong effects. The Monte Carlo standard errors of the mean difference and of the root mean squared difference (x100) were smaller than 0.3 and 0.7, respectively, for all scenarios.
ML, maximum likelihood; FL, Firth's logistic regression; RR, ridge regression.
LOO, leave-one-out crossvalidation; LPO, leave-pair-out crossvalidation; 5-fold, 5-fold crossvalidation; enhBT, enhanced bootstrap; .632+, .632+ bootstrap; app, apparent estimate.

As described in the methods section, LPO CV does not naturally generalize to the Brier score, so we only considered LOO CV, 5-fold CV, enhanced bootstrap and .632+ bootstrap. In all but three simulation scenarios, LOO CV performed best with respect to the mean difference to the IV Brier score for all model estimators, see Figure S2. Similarly as for the discrimination slope, the enhanced bootstrap gave overly pessimistic Brier scores with fairly large RMSD for RR, especially in scenarios with no or small effects. However, differences in RMSD were small between resampling methods, only in scenarios with small or no effects the .632+ bootstrap showed some benefit over the other techniques. With increasing sample size, the RMSD decreased for all resampling techniques and all model estimators.

The percentage of separated data sets was highest (18.2%) for the scenario with a sample size of 50, an event rate of 0.25 and strong effects, see Table S5. In this scenario, more than one third of bootstrap resamples were separated.

## 5. Discussion

Our simulation study does not only confirm that LOO CV yields pessimistic c-statistics [2,3] but also shows that this bias depends on the choice of model estimator. Thus, LOO crossvalidated c-statistics should neither be interpreted as absolute values nor compared between different estimators, e.g. in the optimization of tuning parameters in regularized regression, see Figure S1. LPO CV, which was suggested as an alternative to LOO CV [2], indeed performed better both in terms of mean difference and RMSD to the IV c-statistic. However, the enhanced bootstrap and the .632+ bootstrap achieved a smaller RMSD in almost all simulation settings. The .632+ bootstrap is a weighted average of the apparent c-statistic and a certain overly corrected c-statistic which is replaced by 0.5 if smaller. In this way it is ensured that the .632+ bootstrap c-statistics are greater than or equal to 0.5 (or the apparent c-statistic if smaller). One can apply a similar kind of winsorization with any resampling technique by reporting c-statistics smaller than 0.5 as 0.5. Whereas the practical benefit is questionable, this leads to smaller RMSD to the IV c-statistic in simulations, see Table S6. With this in mind, the superiority of the .632+ bootstrap in terms of RMSD to the IV c-statistic might appear less relevant. Summarizing, we found the performance of LPO CV, 5-fold CV, enhanced bootstrap and .632+ bootstrap in the estimation of c-statistics to be too similar to give definite recommendations, which is in line with a previous study [3]. Thus, the choice might be guided by other criteria such as the dependency on data sampling, the extent of computational burden, the level of complexity of the approach or the likeliness of encountering problems with model fitting due to the sub data structure. If one does not only need to compute the c-statistic but also the receiver operating characteristics curve, one should consider tournament LPO CV [19]. Neither LPO CV nor the bootstrap techniques provide rankings of the data necessary for estimating receiver-operating characteristic curves.

LOO CV cannot be recommended for the estimation of the discrimination slope either, again giving overly pessimistic estimates. Moreover, our simulations revealed unexpected behavior of some of the bootstrap techniques. First, the enhanced bootstrap and the .632+ bootstrap performed reasonably well for ML and FL but sometimes poorly for RR. Second, the simple bootstrap resulted in estimates even more optimistic than the apparent discrimination slopes, see section S3. According to our simulation results we suggest using LPO CV or 5-fold CV to correct for optimism in discrimination slopes.

In general, LOO CV is known to yield nearly unbiased results [1]. However, this result only holds for the averaging approach, where the statistic of interest is applied to each left out observation and then averaged. With the c-statistic and the discrimination slope not applicable to single observations, one has to resort to the pooling approach, i.e. the statistic is calculated only once based on output from different models. The Brier score is an example for a statistic applicable to single observations, where LOO CV can be performed in conjunction with the averaging approach. As expected, the LOO crossvalidated Brier scores were close to unbiased in our simulation.

Our study illustrates that the performance of resampling techniques can vary considerably between model estimators, even if the estimators are similar in construction. Comparing to, e.g., machine learning methods such as support vector machines might even have revealed larger performance differences. This interaction between resampling techniques and model estimators implies that simulation studies aiming to assess the accuracy of a resampling technique should consider a broader set of model estimators to be widely applicable.

## 5. Conclusions

Summarizing, our study emphasizes that estimates provided by resampling techniques should be treated with caution, no matter whether one is interested in absolute values or a comparison between model estimators. Especially in studies with small samples or spurious effects, analysts should not rely on a single resampling technique but should check whether different resampling techniques give consistent results and should report discrepant results.

**Author Contributions:** Conceptualization, A.G. and G.H.; formal analysis, A.G. and G.H.; investigation, A.G., G.H. and L.L.; data curation, M.N.; writing—original draft preparation, A.G.; writing—review and editing, A.G., G.H., L.L. and M.N.; visualization, A.G.; supervision, G.H.; funding acquisition, G.H. All authors have read and agreed to the published version of the manuscript.

**Funding:** This research was funded by the Austrian Science Fund, project number I-2276.

**Acknowledgments:** The data on the screening examination as part of the smoking cessation project was obtained from http://biostat.mc.vanderbilt.edu/DataSets.

**Conflicts of Interest:** The authors declare no conflict of interest. The funders had no role in the design of the study; in the collection, analyses, or interpretation of data; in the writing of the manuscript, or in the decision to publish the results.

## Appendix

The .632+ bootstrap was introduced as a tool providing optimism corrected estimates for error rates [9]. It allows for different choices of the particular form of this error rate but assumes that the error rate can be assessed on the level of observations, i.e. quantifies the discrepancy between a predicted value and the corresponding observed outcome value. As both the c-statistic and the discrimination slope cannot be applied to single observations but only to collections of observations we had to slightly modify the definitions.

The .632+ bootstrap estimate of the c-statistic, $\hat{c}^{.632+}$, is a weighted average of the apparent c-statistic $\hat{c}^{app}$ and an overly corrected bootstrap estimate $\hat{c}^{(1)}$. It is constructed as follows: The model is fitted on each of, say 200 bootstrap resamples (i.e. random samples of size $n$ drawn with replacement) and is used to calculate the predicted probabilities for the observations omitted from the bootstrap resample. For each of the bootstrap resamples the c-statistic is then calculated from the omitted observations. Finally, these c-statistics are averaged over all bootstrap resamples yielding the estimate $\hat{c}^{(1)}$.

The .632+ bootstrap estimate of the c-statistic is then given by

$$\hat{c}^{.632+} = (1 - \hat{w}) \cdot \hat{c}^{app} + \hat{w} \cdot \hat{c}^{(1)},$$

where $\hat{w} = 0.632/(1 - 0.368\,\hat{R})$ with $\hat{R} = (\hat{c}^{app} - \hat{c}^{(1)})/(\hat{c}^{app} - 0.5)$. In order to ensure that $\hat{R}$ falls between 0 and 1 such that $\hat{w}$ ranges from 0.632 to 1, the following modifications are made

- set $\hat{c}^{(1)}$ to 0.5 if $\hat{c}^{(1)}$ is smaller than 0.5 and
- set $\hat{R}$ to 0 if $\hat{c}^{(1)} > \hat{c}^{app}$ or if $0.5 \geq \hat{c}^{app}$.

The value 0.5 occurring in these modifications and in the denominator of $\hat{R}$ is the expected c-index if the outcome is independent of the explanatory variables. The .632+ bootstrap estimate of the discrimination slope can be obtained analogously, just replacing 0.5 by 0 in the definitions above.

Supplement for

# "On resampling methods for model assessment in penalized and unpenalized logistic regression"

by Angelika Geroldinger, Lara Lusa, Mariana Nold, and Georg Heinze

**S1. Problems in resampling techniques associated with small samples**

With small samples one frequently encounters separation in bootstrap resamples or CV subsets even if the original data are not separated. This can lead to problems with methods not being capable of dealing with separated data such as ML or RR. In this study, we decided to follow the simple strategy of restricting the number of iterations in the estimation process and using the results from the last iteration even if ML and RR did not converge due to separation. A more sophisticated strategy would be to replace the model estimation method for separated data subsets by a method that can deal with separation such as Firth's penalization. Another, less frequent problem is the occurrence of bootstrap resamples or CV subsets with linearly dependent explanatory variables, e.g. if a binary explanatory variable is restricted to one category and thus is collinear with the constant. Such a variable would be omitted in a data analysis, but for the sake of simplicity, we just discarded those bootstrap resamples or CV subsets. Finally, the binary outcome might be restricted to one category either in the data subset where the model has to be fitted or in the data subset where the model performance measure is calculated. In both situations, we discarded the affected bootstrap resamples or CV subsets.

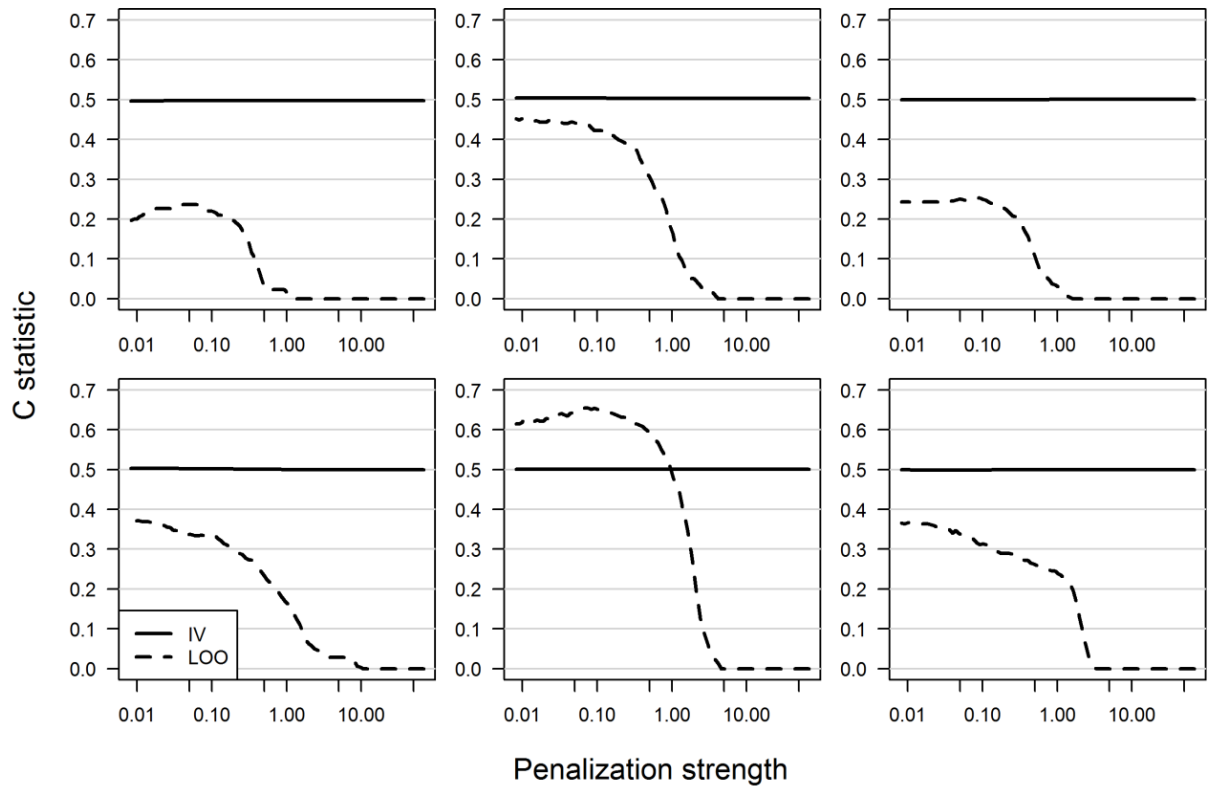

**Figure S1.** Independently validated (solid line) and leave-one-out crossvalidated (dashed line) c-statistics for different penalization strengths in ridge regression on six artificially constructed data sets. The data were created in the same way as for one of the scenarios in our simulation study (null scenario, sample size of 50, marginal event rate of 0.25). The x-axis shows the tuning parameter in ridge regression (lambda in the R package glmnet) with higher values corresponding to stronger penalization. For each data set we fitted 96 ridge regression models corresponding to a series of log-equidistant tuning values. As in our simulation study, the independently validated c-statistics were obtained by validating the models on an independent data set consisting of 100,000 observations. As expected, the independently validated c-statistics are very close to the true value of 0.5.
LOO, leave-one-out crossvalidation; IV, independently validated.

## S2. Data generating mechanism

Data generation was motivated by the structure of real data sets, where typically a mix of variables with different distributions is encountered. We generated one binary ($X_1$), one ordinal ($X_2$) and three continuous ($X_3, X_4, X_5$) explanatory variables as follows: first, we sampled five standard normal deviates $z_{i1}, \dots, z_{i5}$ with correlation matrix as specified in Table S1. Next, we applied the transformations described in the Table S1 to obtain $x_{i1}, \dots x_{i5}$. Finally, we winsorized the continuous variables at the values corresponding to their third quartile plus five times the interquartile distance in each simulated data set. Binary outcomes $y_i$ were drawn from Bernoulli distributions with the event probability following a logistic model, $P(Y|x_{i1}, \dots x_{i5}) = 1/(1 + \exp(-\beta_0 - \beta_1 x_{i1} - \dots - \beta_5 x_{i5}))$. In this way, we obtained data sets with realistic distributions of explanatory variables. For instance, the continuous symmetric variable $X_3$ could represent age and the two right-skewed variables $X_4$ and $X_5$ could represent some lab parameters.

We considered twelve simulation scenarios in a factorial design combining sample size ($n \in \{50, 100\}$), marginal event rate ($E(y) \in \{0.25, 0.5\}$) and effect size (strong or weak effects of all explanatory variables, or null scenarios with no effects). For each scenario we chose the intercept $\beta_0$ such that the desired marginal event rate was approximately achieved. To simulate 'strong effects' scenarios, we set the model coefficients $\beta_1$ to 0.69 and $\beta_2$ to -0.345. For the continuous variables, we set $\beta_3$ to -0.0363, $\beta_4$ to 0.0031, and $\beta_5$ to -0.0039, corresponding to odds ratios of 2 or 1/2 when comparing the fifth and the first sextiles of the distribution functions of the corresponding explanatory variables. To simulate 'weak effects' we set $\beta_1, \dots, \beta_5$ to half of those values. Finally, the null scenarios were obtained by setting $\beta_1, \dots, \beta_5$ to 0. For each scenario we created 1 000 data sets.

**Table S1.** Construction of explanatory variables in the simulation study, following Binder H, Sauerbrei W, Royston P. Multivariable Model-Building with Continuous Covariates: 1. Performance Measures and Simulation Design. Germany: University of Freiburg; 2011. Square brackets [ ... ] indicate that the argument is truncated to the next integer towards 0. The indicator function $\mathbf{1}_{\{\dots\}}$ is equal to 1 if the argument is true and 0 otherwise.

| Underlying variable | Correlation of underlying variables | Explanatory variable | Type | Correlation of explanatory variables |
|---|---|---|---|---|
| $z_{i1}$ | $z_{i3}$ (0.8) | $x_{i1} = \mathbf{1}_{\{z_{i1} < 0.6\}}$ | binary | $x_{i3}(-0.6)$ |
| $z_{i2}$ | $z_{i4}$ (−0.5), $z_{i5}$ (−0.3) | $x_{i2} = \mathbf{1}_{\{z_{i2} \geq -1.2\}} + \mathbf{1}_{\{z_{i2} \geq 0.75\}}$ | ordinal | $x_{i4}$ (−0.4), $x_{i5}$(−0.2) |
| $z_{i3}$ | $z_{i1}$ (0.8) | $x_{i3} = [10 z_{i3} + 55]$ | cont. | $x_{i1}$ (−0.6) |
| $z_{i4}$ | $z_{i2}$ (−0.5), $z_{i5}$ (0.5) | $x_{i4} = [\max(0, 100 \exp(z_{i4}) - 20)]$ | cont. | $x_{i2}(-0.4), x_{i5}(0.4)$ |
| $z_{i5}$ | $z_{i2}$ (−0.3), $z_{i4}$ (0.5) | $x_{i5} = [\max(0, 80 \exp(z_{i5}) - 20)]$ | cont. | $x_{i2}(-0.2), x_{i4}(0.4)$ |

**Table S2.** Mean and standard deviation (x100) of independently validated (IV) c-statistics for different model estimators and all simulation scenarios. The standard deviation strongly depends on the number of new observations (in our case 100 000) used to estimate the IV c-statistics.

| Sample size | Event rate | Effect size | Mean IV c-statistic (x100) | | | Standard deviation of IV c-statistic (x100) | | |
| --- | --- | --- | --- | --- | --- | --- | --- | --- |
| | | | ML | FL | RR | ML | FL | RR |
| 100 | 0.25 | 0 | 49.99 | 49.99 | 49.99 | 0.21 | 0.21 | 0.21 |
| | | 0.5 | 56.87 | 56.83 | 57.15 | 3.43 | 3.44 | 3.6 |
| | | 1 | 67.59 | 67.54 | 67.78 | 2.7 | 2.73 | 2.64 |
| | 0.5 | 0 | 50 | 50 | 50 | 0.18 | 0.18 | 0.18 |
| | | 0.5 | 57.72 | 57.72 | 58 | 2.98 | 2.98 | 3.03 |
| | | 1 | 68.32 | 68.32 | 68.44 | 2.13 | 2.13 | 2.07 |
| 50 | 0.25 | 0 | 50.01 | 50.01 | 50.01 | 0.21 | 0.21 | 0.21 |
| | | 0.5 | 55.38 | 55.26 | 55.63 | 4.14 | 4.21 | 4.48 |
| | | 1 | 64.87 | 64.77 | 65.35 | 4.74 | 4.85 | 4.74 |
| | 0.5 | 0 | 50 | 50 | 50 | 0.19 | 0.19 | 0.19 |
| | | 0.5 | 55.78 | 55.76 | 56.04 | 3.85 | 3.86 | 4.1 |
| | | 1 | 65.78 | 65.76 | 66.24 | 3.92 | 3.93 | 3.63 |

For each simulated dataset and each model estimation method, the IV c-statistic was calculated from newly drawn data with 100,000 observations.

ML, maximum likelihood; FL, Firth's logistic regression; RR, ridge regression.

**Table S3.** Mean and standard deviation (x100) of independently validated (IV) discrimination slope for different model estimators and all simulation scenarios. The standard deviation strongly depends on the number of new observations (in our case 100 000) used to estimate the IV discrimination slope.

| Sample size | Event rate | Effect size | Mean IV discrimination slope (x100) | | | Standard deviation of IV discrimination slope (x100) | | |
|---|---|---|---|---|---|---|---|---|
| | | | ML | FL | RR | ML | FL | RR |
| 100 | 0.25 | 0 | -0.01 | -0.01 | 0 | 0.08 | 0.07 | 0.03 |
| | | 0.5 | 3.27 | 3.05 | 1.73 | 1.97 | 1.84 | 1.65 |
| | | 1 | 11.11 | 10.42 | 8.25 | 3.28 | 3.11 | 3.98 |
| | 0.5 | 0 | 0 | 0 | 0 | 0.07 | 0.07 | 0.03 |
| | | 0.5 | 4.23 | 3.92 | 2.4 | 2.01 | 1.88 | 1.82 |
| | | 1 | 13.49 | 12.63 | 10.73 | 3.11 | 2.99 | 3.89 |
| 50 | 0.25 | 0 | 0 | 0 | 0 | 0.12 | 0.1 | 0.06 |
| | | 0.5 | 3.35 | 2.95 | 1.57 | 2.77 | 2.51 | 2.1 |
| | | 1 | 11.05 | 9.83 | 7.09 | 4.71 | 4.31 | 5.27 |
| | 0.5 | 0 | 0 | 0 | 0 | 0.12 | 0.1 | 0.06 |
| | | 0.5 | 4.03 | 3.53 | 1.94 | 2.92 | 2.61 | 2.29 |
| | | 1 | 13.08 | 11.61 | 8.92 | 4.48 | 4.12 | 5.41 |

For each simulated dataset and each model estimation method, the IV discrimination slope was calculated from newly drawn data with 100,000 observations.

ML, maximum likelihood; FL, Firth's logistic regression; RR, ridge regression.

**Table S4.** Mean and standard deviation (x100) of independently validated (IV) Brier score for different model estimators and all simulation scenarios. The standard deviation strongly depends on the number of new observations (in our case 100 000) used to estimate the IV Brier score.

| Sample size | Event rate | Effect size | Mean IV Brier score (x100) | | | Standard deviation of IV Brier score (x100) | | |
| --- | --- | --- | --- | --- | --- | --- | --- | --- |
| | | | ML | FL | RR | ML | FL | RR |
| 100 | 0.25 | 0 | 20.04 | 19.91 | 19.15 | 0.79 | 0.73 | 0.52 |
| | | 0.5 | 19.41 | 19.28 | 18.88 | 0.81 | 0.75 | 0.53 |
| | | 1 | 17.85 | 17.73 | 17.70 | 0.74 | 0.68 | 0.63 |
| | 0.5 | 0 | 26.66 | 26.42 | 25.52 | 0.94 | 0.83 | 0.61 |
| | | 0.5 | 25.48 | 25.28 | 24.96 | 0.92 | 0.82 | 0.59 |
| | | 1 | 23.01 | 22.87 | 22.88 | 0.86 | 0.78 | 0.76 |
| 50 | 0.25 | 0 | 21.57 | 21.04 | 19.59 | 1.69 | 1.45 | 1.27 |
| | | 0.5 | 20.97 | 20.43 | 19.48 | 1.68 | 1.43 | 1.24 |
| | | 1 | 19.43 | 18.90 | 18.69 | 1.71 | 1.45 | 1.42 |
| | 0.5 | 0 | 28.38 | 27.60 | 26.01 | 1.79 | 1.45 | 1.29 |
| | | 0.5 | 27.31 | 26.59 | 25.76 | 1.95 | 1.62 | 1.42 |
| | | 1 | 24.79 | 24.16 | 24.13 | 1.91 | 1.60 | 1.57 |

For each simulated dataset and each model estimation method, the IV Brier score was calculated from newly drawn data with 100,000 observations.

ML, maximum likelihood; FL, Firth's logistic regression; RR, ridge regression.

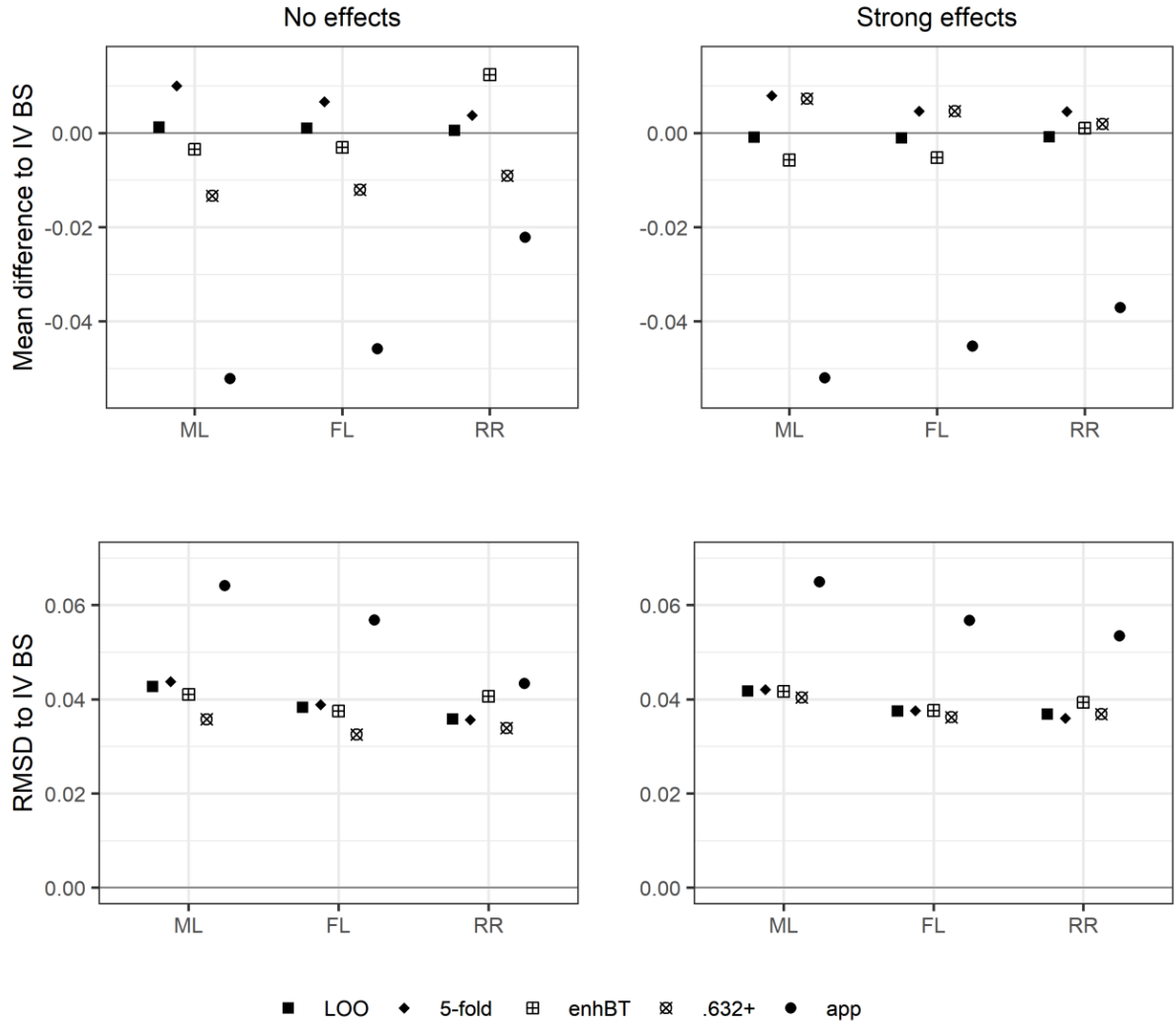

**Figure S2.** Mean and root mean squared differences (RMSD) between Brier scores (BS) computed by data resampling techniques and independently validated (IV) BS for three different model estimators for the simulation settings with 50 observations, an event rate of 0.25 and either no or strong effects. The Monte Carlo standard errors of both, the mean difference and of the root mean squared difference (x100), were smaller than 0.2 for all scenarios.

ML, maximum likelihood; FL, Firth's logistic regression; RR, ridge regression.

LOO, leave-one-out crossvalidation; 5-fold, 5-fold crossvalidation; enhBT, enhanced bootstrap; .632+, .632+ bootstrap; app, apparent estimate.

For each estimator and each resampling technique, the mean difference was calculated as $\frac{1}{1000}\sum_{s=1}^{1000}(b_s - B_s)$, where $b_s$ denotes the Brier score calculated by the respective resampling technique and $B_s$ is the IV Brier score for the respective estimator for the $s$-th generated data set. Its Monte Carlo standard error was obtained as $\left(\frac{1}{999\cdot1000}\sum_{s=1}^{1000}((b_s - B_s) - (\overline{b_s} - \overline{B_s}))^2\right)^{1/2}$, where $\overline{b_s}$ and $\overline{B_s}$ are the mean resampled and IV Brier scores, respectively. The root mean squared difference was computed as $\left(\frac{1}{1000}\sum_{s=1}^{1000}(b_s - B_s)^2\right)^{1/2}$. We used the jackknife to obtain Monte Carlo standard errors of the root mean squared differences, as described in Koehler E, Brown E, Haneuse SJPA. On the Assessment of Monte Carlo Error in Simulation-Based Statistical Analyses. Am Stat. 2009;63(2):155-162.

**Table S5.** Percentage of separated data sets for the twelve simulation scenarios in the full data sets, in the data sets used for model fitting in leave-one-outcrossvalidation, leave-pair-out crossvalidation and 5-fold crossvalidation, respectively, and in the bootstrap data sets.

| Sample size | Event rate | Effect size | Full data | LOO | LPO | 5-fold | BT |
|---|---|---|---|---|---|---|---|
| 100 | 0.25 | 0 | 0.1 | 0.1 | 0.1 | 0.3 | 1.1 |
| | | 0.5 | 0.9 | 0.9 | 0.9 | 1.8 | 4.1 |
| | | 1 | 2.6 | 2.7 | 2.9 | 5.8 | 10.8 |
| | 0.5 | 0 | 0.0 | 0.0 | 0.0 | 0.0 | 0.0 |
| | | 0.5 | 0.0 | 0.0 | 0.0 | 0.0 | 0.1 |
| | | 1 | 0.0 | 0.0 | 0.0 | 0.1 | 0.4 |
| 50 | 0.25 | 0 | 2.9 | 3.1 | 3.6 | 6.3 | 12.5 |
| | | 0.5 | 9.5 | 9.9 | 10.3 | 15.1 | 23.3 |
| | | 1 | 18.2 | 18.8 | 20.0 | 26.0 | 35.9 |
| | 0.5 | 0 | 0.1 | 0.1 | 0.2 | 0.7 | 2.6 |
| | | 0.5 | 0.3 | 0.3 | 0.4 | 1.1 | 3.6 |
| | | 1 | 1.2 | 1.3 | 1.4 | 2.9 | 6.9 |

We checked for separation using Ioannis Kosmidis' R package brglm2, version 0.1.8.

LOO, leave-one-out crossvalidation; LPO, leave-pair-out crossvalidation; 5-fold, 5-fold crossvalidation; BT, bootstrap.

**Table S6.** Mean difference and root mean squared difference (x100) between winsorized c-statistics computed by different resampling techniques and the independently validated (IV) value (as presented in Table S2) for simulation scenarios with sample size of 50 and event rate of 0.25. Resampled c-statistics were winsorized by replacing values smaller than 0.5 by 0.5. Figure 3 shows the analogous results for the untransformed c-statistics.

| Effect size | Estimator | Mean difference (x100) | | | | | | Root mean squared difference (x100) | | | | | |
|---|---|---|---|---|---|---|---|---|---|---|---|---|---|
| | | LOO | LPO | 5-fold | enhBT | .632+ | app | LOO | LPO | 5-fold | enhBT | .632+ | app |
| 0 | ML | 2.76 | 5.53 | 4.99 | 8.08 | 5.08 | 20.20 | 5.98 | 9.21 | 8.53 | 11.26 | 8.89 | 21.48 |
| | FL | 2.87 | 5.61 | 5.08 | 8.10 | 5.10 | 20.06 | 6.29 | 9.29 | 8.63 | 11.30 | 8.91 | 21.34 |
| | RR | 1.45 | 5.68 | 5.13 | 7.25 | 5.05 | 18.47 | 4.73 | 9.34 | 8.74 | 10.67 | 8.89 | 19.91 |
| 0.5 | ML | -0.07 | 3.36 | 2.68 | 6.02 | 3.08 | 17.50 | 7.91 | 9.68 | 9.17 | 10.83 | 9.57 | 19.12 |
| | FL | 0.02 | 3.40 | 2.65 | 6.00 | 2.87 | 17.46 | 8.01 | 9.68 | 9.15 | 10.80 | 9.45 | 19.08 |
| | RR | -2.07 | 3.32 | 2.60 | 5.00 | 2.72 | 15.71 | 8.21 | 9.81 | 9.24 | 10.51 | 9.56 | 17.69 |
| 1 | ML | -3.51 | 0.96 | 0.01 | 3.30 | 0.61 | 13.17 | 10.34 | 10.21 | 9.99 | 10.20 | 10.36 | 15.35 |
| | FL | -3.26 | 0.98 | -0.02 | 3.22 | 0.33 | 13.12 | 10.40 | 10.21 | 9.98 | 10.18 | 10.31 | 15.32 |
| | RR | -6.02 | 0.70 | -0.17 | 2.40 | 0.21 | 11.69 | 11.92 | 10.17 | 9.92 | 10.28 | 10.35 | 14.38 |

ML, maximum likelihood; FL, Firth's logistic regression; RR, ridge regression.

LOO, leave-one-out crossvalidation; LPO, leave-pair-out crossvalidation; 5-fold, 5-fold crossvalidation; enhBT, enhanced bootstrap; .632+, .632+ bootstrap; app, apparent estimate.

For each estimator and each resampling technique, the mean difference was calculated as $\frac{1}{1000}\sum_{s=1}^{1000}(c_s - C_s)$, where $c_s$ denotes the winsorized c-statistic calculated by the respective resampling technique and $C_s$ is the IV c-statistic for the respective estimator for the s-th generated data set. The root mean squared difference was computed as $\left(\frac{1}{1000}\sum_{s=1}^{1000}(c_s - C_s)^2\right)^{1/2}$.

**S3. A side remark on the simple bootstrap: resampling may increase the optimism**

With the simple bootstrap, the parameter estimates from models fitted on bootstrap resamples (sampling $n$ observations with replacement from the original data) are used to calculate the c-statistic for the original data sample, see *Efron B, Tibshirani RJ. An Introduction to the Bootstrap: Chapman & Hall; 1993*. Usually this is repeated, say, 200 times and the estimates are averaged. The simple bootstrap is known to perform poorly compared to the more refined bootstrap techniques which we also considered in our study, see *Harrell FE. Regression Modeling Strategies: Springer; 2001*.

We have not included the simple bootstrap in the main presentation of our simulation results due to its known inferiority. Though, some results are worth to report. In all simulation scenarios, the simple bootstrap gave median discrimination slopes even more optimistic than the apparent ones. In other words, the simple bootstrap increased the optimism instead of correcting it as we would expect. This phenomenon was observed with each model estimator. At first glance, it might appear counterintuitive that models fitted on bootstrap resamples discriminate the original outcomes better than the model fitted on the original data, but there is a simple explanation: models fitted on bootstrap resamples with their repeated observations tend to give more extreme predicted probabilities than the model fitted on the original data.

The c-statistics and Brier scores estimated by the simple bootstrap were also severely overoptimistic but on average smaller than their apparent counterparts.